\documentclass[onecolumn,showpacs,showkeys,preprintnumbers,amsmath,amssymb]{revtex4}

\usepackage{amsmath, amssymb, latexsym, verbatim}
\usepackage{graphicx}
\usepackage{dcolumn}
\usepackage{bm}
\usepackage[dvips]{color}
\usepackage[usenames,dvipsnames]{xcolor}

\newcommand{\nin}{\noindent}

\begin{document}

\title{A Quasi-Newtonian Approach to Bohmian Mechanics II: Inherent Quantization}

\author{Mahdi Atiq}
 \altaffiliation{Sharif University of Technology, Physics Department, Tehran, Iran}
 \email{mma\_atiq@yahoo.com}

\author{Mozafar Karamian}
 \email{karamian@ymail.com}

\author{Mehdi Golshani}
 \altaffiliation{Institutes for Theoretical Physics and Mathematics (IPM), Tehran, Iran}
 \altaffiliation{Department of Physics, Sharif University of Technology, Tehran,
        Iran}
 \email{mehdigolshani@yahoo.com}

\begin{abstract}
\nin In a previous paper, we obtained the functional form of
quantum potential by a quasi-Newtonian approach and without
appealing to the wave function. We also described briefly the
characteristics of this approach to the Bohmian mechanics. In this
article, we consider the quantization problem and we show that the
'eigenvalue postulate' is a natural consequence of continuity
condition and there is no need for postulating that the spectrum
of energy and angular momentum are eigenvalues of their relevant
operators. In other words, the Bohmian mechanics \emph{predicts}
the 'eigenvalue postulate'.
\end{abstract}



\maketitle

\emph{Published in Annales de la Fondation Louis de Broglie, 2009}

\section{Introduction} \label{S:Intro}
\nin In a previous paper, we considered a quasi-Newtonian approach
to the Bohmian mechanics. We could obtain the functional form of
the quantum potential without appealing to any wave function. In
this article, we consider the method of solving quantum problems
in this quasi-Newtonian approach and we show that the 'eigenvalue
postulate' for energy and angular momentum are natural consequence
of the continuity equation and there is no need for postulating it
in the Bohmian mechanics. This fact throws light on the problem of
eigenvalues in the quantum theory. There is a hope that one can
generalize this statement for physical quantities other than
energy and orbital angular momentum, specially for \emph{spin}.

In the previous paper, we also described the main differences
between our quasi-Newtonian approach to Bohmian mechanics and the
usual one. We mentioned that the imposition of the uniqueness
condition on $S$ and positive-definiteness condition on $R$ are
not necessary for solving quantum problems. In this article, we
practically see these facts by solving some specific problems.

\section{Solving problems in the quasi-Newtonian approach to Bohmian mechanics}
\label{S:New Approach}

\subsection{About the eigenvalue postulate} \label{S:The eigenvalue post.}
\nin Quantum mechanics consists of two basic parts: Schr\"odinger
equation and operators. On the one hand, we assume that the wave
function of any physical system satisfies the time-dependent
Schr\"odinger equation, and on the other hand, for any physical
quantity like energy, momentum, etc., there exists an operator,
the eigenvalues of which constitute all possible values of that
quantity. This assumption does not result from the Schr\"odinger
equation, and in a certain sense is one of the important
postulates of physics, since Galileo and Newton. The
\emph{eigenvalue postulate} raises some questions. For example,
all operators do not have common eigenfunctions (like e.g. the
components of angular momentum operator). What can one do with
physical quantities corresponding to these operators? How can one
get their values and how are they defined? If we take this
postulate seriously, then one must conclude that all of the
components of angular momentum do not exist at the same time. Then
the question arises about the physical meaning of the
\emph{non-existence} of a component of angular momentum, or how
can one imagine a physical quantity without having any definite
value? Any alternative theory that tries to explain quantum
phenomena, without accepting this postulate, has to explain why
the measured values of the physical quantities are simply the
eigenvalues of their corresponding operators? In the Bohmian
interpretation of quantum mechanics, it is possible to show how
the process of measurement terminates with an eigenvalue of the
relevant quantum operator without any collapse of the wave
function \cite{Bo52},\cite[Chap. 8]{Ho93}. Of course, it seems
that the concept of operators and operator relations and their
roles in the quantum theory are not made clearer
in the ordinary Bohmian interpretation of quantum mechanics.\\

When we want to find the energy levels of atoms in the Bohmian
mechanics, we solve the time-independent Schr\"odinger equation

\begin{equation}\label{E:H psi = E psi}
    \hat{H} \psi = E \psi
\end{equation}

\noindent as in the standard quantum mechanics. Also, for finding
angular momentum values we solve the equations

\begin{equation}\label{E:L2 psi = l(l+1) hbar2}
    \hat{L}^2 \psi = l(l+1) \hbar^2 \psi
\end{equation}

\begin{equation}\label{E:Lz psi = m hbar}
    \hat{l}_z \psi = m \hbar \psi.
\end{equation}

These equations are postulates of the standard quantum mechanics.
Why we must use them in the Bohmian mechanics too? In the Bohm's
postulates there are no operators and operator relations.
Therefore, we must \emph{prove} these relations in the Bohmian
mechanics.

\indent In the next sections of this paper, we attempt to show
that one can solve quantum mechanical problems without starting
from operator methods. Also, one can clarify the meaning of
operator methods, specially eigenvalue problems. We show that one
can \emph{prove} the validity of eigenvalue equations, and during
this proof the meaning of these equations becomes clearer.
Consequently, we shall discuss about the situations that we can
use or not use these equations.

\subsection{The stationary states as conservative states} \label{S:Stationary States}
\nin In classical mechanics, whenever classical potential $V$ does
not depend on time, the 'system' is conservative. For example,
when you throw a particle with an arbitrary velocity, the energy
of system remains invariant ($dH/dt=\partial V /
\partial t = 0$). But, in Bohmian mechanics, because of the presence
of quantum potential $Q$, we have

\begin{equation}\label{E:dH/dt}
    dH/dt=\partial (V+Q) / \partial t = \partial Q / \partial t.
\end{equation}

Therefore, for the motion to be conservative, we must have

\begin{equation}\label{E:Conserv. Condition}
    \partial Q / \partial t = 0.
\end{equation}

When $R$ tends to zero at infinity (or at system boundary), this
equation reduces to $\partial R / \partial t = 0$, i.e., the state
must be stationary. Therefore, in stationary states the motion is
conservative, and due to this fact we can call the stationary
states as 'conservative states'. Thus, in the Bohmian mechanics,
even when the potential $V$ is classically conservative, the
energy is not necessarily conservative. The conservative motions
are restricted to the cases that the state is stationary. This is
the reason for the significance of stationary states, among
arbitrary states.

\subsection{Hamilton's canonical equations and quantization problem} \label{S:Canonical Eq
Approach} \nin The central concept in the ordinary quantum
mechanics, is that of wave function. In the case of stationary
states (where $\partial R / \partial t=0$ and $S=W(x)-Et$ ), we
look for the eigenfunctions and the eigenvalues of the operators
(accurately or approximately). In the scattering problems as well
as those problems involving time-dependent potentials, the wave
function plays a central role. But a question arises about whether
the concept of wave function is necessary for the conceptual
structure and the mathematical formulation of the quantum
mechanics. Is it really necessary to consider both $R$ and $S$ as
physical functions which are unique at all points of space (apart
from additive constants for $S$)? Is it not sufficient to refer
quantum phenomena only to $R$ and regard $S$ as an auxiliary
(multi-valued) function like its classical counterpart, and not as
the phase of a wave function? To what extent appealing to wave
function and operator-based approaches is necessary for solving
quantum problems?\\
\indent One possibility for finding the answer of these questions
is to add the quantum potential to the classical potential and try
to solve the problems through Hamilton's canonical equations to
describe the quantum phenomena like interference, passage through
a potential barrier, $\cdots$ , and specifically the quantization
of physical quantities (energy, angular momentum, $\cdots$). What
that justifies this method is that the quantum potential appears
in the Hamiltonian as a function of space. Thus, its role in
Hamilton's canonical equations is similar to the role of classical
potential. But, one must note that the mere addition of a
function, named quantum potential, to the canonical equations,
does not bring about the quantization of the quantities, as it
doesn't occur for any classical potential. The cause of
quantization lies elsewhere. We shall see how, by adding quantum
potential to the mechanics of the particle and regarding the
continuity equation, the quantization of quantities (energy and
angular momentum, in this article) for bound conserved states
\emph{becomes a necessity}. We show that the role of continuity
equation in the quantization of quantities is essential. We prove
that, the continuity condition is sufficient for understanding the
fact that the energy and angular momentum of particle in the
stationary states are simply the eigenvalues of the relevant
operators and there is no need to take this statement as a
postulate. This fact shows a merit of Bohmian views (in the
framework of a quasi-Newtonian approach) that reduces the number
of postulates we need to explain quantum phenomena.\\
\indent In order to show that this method works, we shall solve
some problems by this method in the next sections and show that in
order to solve quantum problems, it does not seem necessary to use
the usual \emph{operator-based} methods used in the standard
quantum mechanics. We shall show that one can get
\emph{quantization without any reference to the wave function and
the eigenvalue postulate}.

\section{The analysis of some problems using canonical equations}
\label{S:Analysis of some problems}

\subsection{Stationary (conservative) states in one dimension} \label{S:Stationary states 1D}
\nin Consider a one-dimensional single-particle system with the
Hamiltonian:

\begin{equation} \label{E:1D Hamiltonian}
H(x, p_x)=\frac{p_x^2}{2m}+V(x)+Q(x)
\end{equation}

\noindent in which the quantum potential $Q$  is of the form:

\begin{equation} \label{E:1D Q}
Q(x)=-\frac{\hbar^2}{2m}\frac{1}{R}\frac{d^2 R}{dx^2}, \quad
R=R(x).
\end{equation}

The canonical equations are

\begin{align}
\dot{x}&=\frac{\partial H}{\partial p_x}=\frac{p_x}{m} \nonumber \\
\dot{p}_x &=-\frac{\partial H}{\partial
x}=-\frac{dV}{dx}-\frac{dQ}{dx}. \nonumber
\end{align}

\noindent From these two equations, one concludes that

\begin{align}
p_x^2+2m(V+Q)=\alpha \nonumber
\end{align}

\noindent where $\alpha$ is the constant of integration, and by
equating $H$ with $E$, one gets $\alpha=2mE$. Therefore, for a
stationary state in one-dimensional potentials, we have

\begin{equation} \label{E:1D QHJ}
\frac{R}{\hbar^2}p_x^2=\frac{d^2R}{dx^2}+2m(E-V(x))\frac{R}{\hbar^2}.
\end{equation}

There is another important condition that must be considered which
is the continuity equation

\begin{equation} \label{E:1D Contin -Stationary}
\frac{d}{dx}(R^2 p_x)=0.
\end{equation}

In classical mechanics, one obtains the classical laws of motion,
by taking the extremum of an integral of action along the path of
the particle, whereas the continuity equation is obtained by
identifying the canonical distribution function $f(x, p_x)$ with
$R^2(x) \delta(p_x - \partial S /
\partial x)$ and using Liouville's equation \cite[Chap.
2]{Ho93}. Therefore, in classical mechanic the expression
$d^2R/dx^2$ does not enter into the equation
\eqref{E:1D QHJ}, without any need to have $d^2R/dx^2=0$ .\\
\indent Now, we pay attention to the condition \eqref{E:1D Contin
-Stationary} which implies that $R^2p_x=\lambda=const$. The
constant $\lambda$ is the same throughout the particle path, being
either zero or non-zero. If $\lambda \neq 0$ , then we should
always have $p_x \neq 0$ because $R$ as physical quantity can not
be infinite, which means that the particle never stops and thus
cannot have classical turning points. This state is not a bound
one. But, if $\lambda=0$, then because $R$ is not always zero, we
must have  $p_x=0$. Thus, the particle would be at rest.
Therefore, a bound particle in a stationary state is always at
rest. Now, the question arises: whereas the continuity equation is
independent of the form of quantum potential $Q$  as a function of
$R$ and the continuity equation is common between classical and
quantum mechanics, why we do have turning points in classical
mechanics for bound particles? In response, one may suggest that
the classical dynamics must be considered as non-stationary states
with $R=R(x, t)$  and $S=W(x)-Et$. Therefore, we must consider

\begin{equation} \label{E:1D Contin Eq}
\frac{\partial}{\partial t}R^2+\frac{\partial}{\partial x} ( R^2
\frac{p_x}{m} )=0
\end{equation}

\noindent instead of equation \eqref{E:1D Contin -Stationary} for
classical dynamics. Therefore, because $R$ in classical mechanics
appears only in the continuity equation \eqref{E:1D Contin Eq} and
not in the energy equation, one can always find a well-behavior
function $R(x, t)$ such that it satisfies the equation \eqref{E:1D
Contin Eq}.
Therefore, turning point \emph{becomes possible}.\\
\indent Thus, in the bound stationary states of quantum mechanics,
the particle is at rest and the energy eigenvalues are obtained
from

\begin{equation} \label{E:1D R eigenvalues equation}
\frac{d^2 R}{dx^2}+\frac{2m}{\hbar^2}(E-V(x))R=0.
\end{equation}

This is the time-independent Schr\"odinger equation for a real
time-independent wave function $R$. In the standard quantum
mechanics, it is only after the introduction of the time-reversal
operator and the imposition of the invariance under this operator
that we prove that in the case of one-dimensional bound states,
the wave function is real and
therefore $R$ replace $\psi$  in the eigenvalue relation.\\

\indent An important point about the equation \eqref{E:1D R
eigenvalues equation} is that its solutions for $R$  are not
necessarily positive definite functions. This is a fact that we
mentioned in the section \ref{S:Intro}. For example, for the case
of a particle in a box, we have

\begin{equation*}
V(x)= \left\{%
\begin{array}{ll}
    0 & \text{if \,} 0 < x < a  \\
    \infty & \text{otherwise.} \\
\end{array}%
\right.
\end{equation*}

\noindent The solutions are

\begin{equation*}
    R(x) \propto \sin \frac{n \pi x}{a}
\end{equation*}

\noindent in which  $R$ is negative in certain intervals. Thus the
$R$ that appears here is not exactly the same as $R$ that appears
in Bohm's formulation. If we write $R$  as

\begin{equation*}
    R= \left| R \right| e^{i \chi}
\end{equation*}

\noindent then, for $R \geq 0$, we have $\chi = 0$  and for $R <
0$ , we have $\chi = \pi$. In the usual formulation of Bohmian
mechanics, $\chi$ is a part of the phase of the wave function.
Thus the phase of wave function is equal to

\begin{equation*}
    S^\prime(x, t)/\hbar=\chi-Et/\hbar.
\end{equation*}

\noindent We have denoted the phase of wave function by
$S^\prime$, because in our approach here $S=-Et/\hbar$. In the
usual Bohmian mechanics, although $S^\prime$ is not the same at
all points, but as it is constant in any interval between the zero
points of $R$, the particle remains at rest. In this situation,
the value of $S^\prime$ near the points where $R$ is equal to
zero, i.e., at nodes, changes non-continuously. But, from the
point of view adopted in this paper, $S$ is continuous. The
function $S$ is constant throughout the whole box and the value of
$R$ changes continuously between positive and negative values.
Therefore, it is more suitable to let $R$ take negative values
too. Thus, its interpretation as the amplitude of a wave is not
correct, because the amplitude of a wave cannot be negative. Of
course, in the Bohmian mechanics, $R$  appears in the form of
$R^2$ or $R^{-1} \nabla^2 R$. Thus, negative $R$ is not a problem,
and the condition $R \geq 0$ does not seem to be necessary. This
supports our argument that appealing to the wave function is not
necessary for quantum theory. As we shall see in the case of
central potential in three dimension too, the positive
definiteness of $R$ is superfluous.

\subsection{Stationary (conservative) states of central potentials} \label{S:Central potentials}
\nin Consider the case of a particle which moves in a three
dimensional central problem with a Hamiltonian

\begin{equation}\label{E:3D Hamiltonian -polar}
    H(r, \theta, \phi, p_r, p_{\theta}, p_{\phi})= \frac{1}{2m}( p_r^2+ \frac{p_{\theta}^2}{r^2}+ \frac{p_{\phi}^2}{r^2 \sin^2 \theta}
    ) + V(r)+ Q(r, \theta, \phi).
\end{equation}

Now, considering the spherical symmetry of $V(r)$, we write $R$ as

\begin{equation}\label{E:R -polar}
    R(r, \theta, \phi)=R_r(r)R_{\theta}(\theta)R_{\phi}(\phi).
\end{equation}

\noindent Then the quantum potential takes the form

\begin{equation}\label{E:Q -polar}
    Q(r, \theta, \phi)=Q_r(r)+\frac{Q_{\theta}(\theta)}{r^2}+\frac{Q_{\phi}(\phi)}{r^2 \sin^2 \theta}
\end{equation}

\noindent in which

\begin{equation}\label{E:Qr}
    Q_r(r)=-\frac{\hbar^2}{2m}\frac{1}{R_r}\frac{1}{r^2}\frac{\partial}{\partial
    r}( r^2 \frac{\partial}{\partial r} R_r )
\end{equation}

\begin{equation}\label{E:Qtheta}
    Q_{\theta}(\theta)=-\frac{\hbar^2}{2m}\frac{1}{R_{\theta}}\frac{1}{\sin \theta}\frac{\partial}{\partial
    \theta}( \sin \theta \frac{\partial}{\partial \theta} R_{\theta} )
\end{equation}

\begin{equation}\label{E:Qphi}
    Q_{\phi}(\phi)=-\frac{\hbar^2}{2m}\frac{1}{R_{\phi}}\frac{\partial^2}{\partial
    \phi^2} R_{\phi}.
\end{equation}

The canonical equations of motion for the $\phi$  coordinate are

\begin{align*}
    \dot{\phi}&=\frac{\partial H}{\partial p_\phi}=\frac{p_\phi}{mr^2 \sin^2
    \theta}\\
    \dot{p}_\phi&=-\frac{\partial H}{\partial \phi}=-\frac{1}{r^2 sin^2
    \theta}\frac{\partial Q_\phi}{\partial \phi}\\
\end{align*}

\noindent which lead to

\begin{equation}\label{E:p_phi2}
    p_\phi^2+2mQ_\phi(\phi)=\alpha_\phi^2
\end{equation}

\noindent in which $\alpha_\phi^2$ is the constant of integration
and thus a constant of motion. Similarly, the canonical equations
for the coordinate $\theta$ and $r$ lead to

\begin{equation}\label{E:p_theta2}
    p_\theta^2+\frac{\alpha_\phi^2}{\sin^2
    \theta}+2mQ_\theta(\theta)=\alpha_\theta^2
\end{equation}

\begin{equation}\label{E:pr2}
    H=\frac{1}{2m}( p_r^2+\frac{\alpha_\theta^2}{r^2}
    )+V(r)+Q_r(r)=E.
\end{equation}

The equations \eqref{E:p_phi2}-\eqref{E:pr2} can be written in the
form:

\begin{equation}\label{E:p_phi2 Main Eq}
    \frac{R_\phi}{\hbar^2}p_\phi^2=\frac{\partial^2}{\partial
    \phi^2}R_\phi+\frac{\alpha_\phi^2}{\hbar^2}R_\phi
\end{equation}

\begin{equation}\label{E:p_theta2 Main Eq}
    \frac{R_\theta}{\hbar^2}p_\theta^2=\frac{1}{\sin
    \theta}\frac{\partial}{\partial \theta}( \sin \theta \frac{\partial}{\partial \theta}R_\theta )
    +( \alpha_\theta^2-\frac{\alpha_\phi^2}{\sin^2 \theta}
    )\frac{R_\theta}{\hbar^2}
\end{equation}

\begin{equation}\label{E:pr2 Main Eq}
    \frac{R_r}{\hbar^2}p_r^2=\frac{1}{r^2}\frac{\partial}{\partial
    r}( r^2 \frac{\partial}{\partial r}R_r
    )+\frac{2m}{\hbar^2} \left[ E-V(r)-\frac{\alpha_\theta^2}{2mr^2}
    \right] R_r.
\end{equation}

If the left-hand sides of the equations \eqref{E:p_theta2 Main Eq}
and \eqref{E:pr2 Main Eq} are zero, then these two equations
reduce to the angular and radial Schr\"odinger equations. For the
case of equation \eqref{E:p_phi2 Main Eq} and the quantization of
$\alpha_\phi^2$ , we shall talk about later. We observe that, for
$\alpha_\theta^2$ and $E$ to be quantized, it is necessary for
$p_\theta$ and $p_r$ to be zero, and we shall see that the
continuity condition compel these quantities to be zero.

In the case of one dimensional problem, we saw that it was more
natural to let $R$ taking negative values as well. Here, as it is
clear from the relevant equations, the functions $R_r$, $R_\theta$
and $R_\phi$ can take negative values too. Thus, again, the
interpretation of $R$ as the amplitude of the wave function is not suitable.\\

\indent What is the meaning of the constants $\alpha_\theta$  and
$\alpha_\phi$? In order to see meaning of these constants, it
would be better to compare the equations \eqref{E:p_phi2} to
\eqref{E:pr2} with their classical counterparts. In the classical
case\cite[pp. 450-451]{Go02}, we have

\begin{equation}\label{E:Classic p_phi Eq}
    p_\phi=\alpha_\phi
\end{equation}

\begin{equation}\label{E:Classic p_theta Eq}
    p_\theta^2+\frac{\alpha_\phi^2}{\sin^2
    \theta}=\alpha_\theta^2=l^2
\end{equation}

\begin{equation}\label{E:Classic pr Eq}
    p_r^2+\frac{\alpha_\theta^2}{r^2}+2mV(r)=2mE
\end{equation}

\noindent where $l$ is the magnitude of the classical angular
momentum. In the classical case, $\alpha_\phi$  is the
$z$-component of angular momentum and $\alpha_\theta$ is the
magnitude of the angular momentum. But, in the quantum case, even
though $\alpha_\theta$  and $\alpha_\phi$ are constants, they do
not have exactly the same meaning. Here the $z$-component of
angular momentum $p_\phi$ and the magnitude of the angular
momentum vector ${( p_\theta^2 + {p_\phi^2}/{\sin^2 \theta}
)}^{1/2}$ are not necessarily constants of motion. What we measure
in the laboratory as the $z$-component of angular momentum
$(m\hbar)$ and the square of the magnitude of the angular momentum
vector $(l(l+1)\hbar^2)$, are in fact the constants $\alpha_\phi$
and $\alpha_\theta^2$, and not the real values of the
$z$-component of angular momentum and the square of the magnitude
of the angular momentum vector respectively. Now, we discuss about
the cause of quantization of these quantities.\\

Before continuing, we consider a lemma from classical
Hamilton-Jacobi theory. We know that when the classical potential
is in the form of

\begin{equation}\label{E:Full form of V}
    V(r, \theta, \phi) = V_r(r)+ \frac{V_\theta(\theta)}{r^2}+ \frac{V_\phi(\phi)}{r^2 sin^2 \theta}
\end{equation}

\noindent then, the principal function $S$, becomes fully
separable, in the form of

\begin{equation}\label{E:S=Wr+Wt+Wp-Et}
    S(r, \theta, \phi) = W_r(r) + W_\theta(\theta) + W_\phi(\phi)
    - Et.
\end{equation}

This is the case that we have here, because, considering
\eqref{E:Q -polar}, the effective potential $V+Q$ is exactly in
the form of \eqref{E:Full form of V}. Therefore, the equation
\eqref{E:S=Wr+Wt+Wp-Et} is valid for our problem.

\indent Similar to the one dimensional problem in previous
section, here we must discuss about the condition of continuity.
But, to begin with, it would be better to pay attention to several
mathematical identities that clarify the relation between the
continuity equation and the eigenvalue equations of quantum
operators. First, we define the following expressions

\begin{equation}\label{E:fr}
    f_r(r)=\frac{1}{R_r^2}\frac{\partial}{\partial r}(
    r^2 R_r^2 \frac{\partial W_r}{\partial r} )
\end{equation}

\begin{equation}\label{E:f theta}
    f_\theta(\theta)=\frac{1}{R_\theta^2}\frac{1}{\sin \theta}\frac{\partial}{\partial
    \theta}( \sin \theta R_\theta^2 \frac{\partial W_\theta}{\partial \theta} )
\end{equation}

\begin{equation}\label{E:f phi}
    f_\phi(\phi)=\frac{1}{R_\phi^2}\frac{\partial}{\partial
    \phi}( R_\phi^2 \frac{\partial W_\phi}{\partial \phi} ).
\end{equation}

By these definitions the continuity condition becomes

\begin{equation}\label{E:Contin Eq Expaned}
    0=\frac{r^2}{R^2} \nabla . ( R^2 \nabla S ) = f_r+f_\theta+ \frac{f_\phi}{\sin^2 \theta}.
\end{equation}

Writing $\psi$  in the form $\psi=R \, e^{iS/\hbar}$ we get, after
some algebra

\begin{equation}\label{E:Lz psi on psi}
    \frac{\hat{L}_z
    \psi}{\psi}= {p_\phi} + (-i\hbar) \frac{1}{R_\phi}\frac{\partial R_\phi}{\partial \phi}
\end{equation}

\begin{equation}\label{E:Lz2 psi on psi}
    \frac{\hat{L}_z^2 \psi}{\psi}= \left( p_\phi^2 + 2mQ_\phi \right)+ (-i\hbar)f_\phi = \alpha_\phi^2+(-i\hbar)f_\phi
\end{equation}

\begin{equation}\label{E:L2 psi on psi}
    \frac{\hat{L}^2 \psi}{\psi}
    =  \left( p_\theta^2 + \frac{p_\phi^2}{\sin^2 \theta} \right) + 2m \left( Q_\theta+\frac{Q_\phi}{\sin^2 \theta}
    \right)  + (-i\hbar) \left(  f_\theta+ \frac{f_\phi}{\sin^2 \theta} \right) \\
    = \alpha_\theta^2 + (-i\hbar) \left(  f_\theta+ \frac{f_\phi}{\sin^2 \theta} \right)
\end{equation}

\begin{equation}\label{E:H psi on psi}
    \frac{\hat{H} \psi}{\psi}= \Big\{ \frac{( \nabla S )^2}{2m} +V+Q
    \Big\} + \frac{(-i\hbar)}{2R^2}\nabla . ( R^2 \frac{\nabla S}{m}
    )=E.
\end{equation}

We see that, $E$ is eigenvalue of $\hat{H}$ (because of
\eqref{E:Contin Eq Expaned}, the imaginary part of \eqref{E:H psi
on psi} is zero), but the other constants of motion
$\alpha_\theta^2$ and $\alpha_\phi^2$ are not necessarily
eigenvalues of $\hat{L}^2$ and $\hat{L}_z^2$, respectively. Also,
$p_\phi$ is not eigenvalue of $L_z$, necessarily.

If we want $\psi$ to be the simultaneous eigenfunction of
operators $\hat{L}_z^2$, $\hat{L}^2$ and $\hat{H}$, the different
parts of the continuity equation must be separately zero, i.e.,

\begin{equation}\label{E:f r theta phi = 0}
    f_\phi = 0, \, f_\theta = 0, \, f_r = 0.
\end{equation}

In other words, the validity of the Eqs. \eqref{E:f r theta phi =
0} is equivalent to $\psi$ being the simultaneous eigenfunction of
$\hat{L}_z^2$ , $\hat{L}^2$ and $\hat{H}$ (in a special case
$\psi$ may be the eigenfunction of $\hat{L}_z$).

But from equation \eqref{E:Contin Eq Expaned}, we conclude that
for some constants $c_\phi$ and $c_\theta$ we have

\begin{equation}\label{E:f phi=const}
    f_\phi(\phi)=c_\phi
\end{equation}

\begin{equation}\label{E:f theta=const}
    f_\theta(\theta)+\frac{c_\phi}{\sin^2 \theta}=c_\theta
\end{equation}

\begin{equation}\label{E:fr=const}
    f_r(r)=-c_\theta
\end{equation}

\noindent and from Eqs. \eqref{E:Lz2 psi on psi} and \eqref{E:L2
psi on psi} we get

\begin{equation}\label{E:Lz2 psi on psi -2}
    \frac{\hat{L}_z^2 \psi}{\psi}=\alpha_\phi^2-i\hbar c_\phi
\end{equation}

\begin{equation}\label{E:L2 psi on psi -2}
    \frac{\hat{L}^2 \psi}{\psi}=\alpha_\theta^2-i\hbar c_\theta.
\end{equation}

When the constants $c_\theta$  and $c_\phi$  are not necessarily
zero, $\psi$ is not necessarily an eigenfunction of $\hat{L}_z^2$
and $\hat{L}^2$, whereas $\alpha_\phi^2$ and $\alpha_\theta^2$ are
still constants of motion. Now, we consider the values of
constants
$c_\phi$, $c_\theta$ for a bound system.\\

When we have identically $\dot{r} = 0$, we get from Eq.
\eqref{E:fr=const} that $c_\theta = 0$. If the orbit of particle
is such that $\dot{r}$ is not equal to zero at all points, there
necessarily exist at least one $r_{min}$ and one $r_{max}$ in the
orbit of particle, for the system to be bound. If we integrate Eq.
\eqref{E:fr=const} from one $r_{min}$ to the subsequent $r_{max}$,
we obtain

\begin{equation}\label{E:fr integrate}
    (r^2 R_r^2 \frac{\partial W_r}{\partial r})_{r_{max}} - (r^2 R_r^2 \frac{\partial W_r}{\partial r})_{r_{min}}
    = - c_{\theta} \int_{r_{min}}^{r_{max}} R_r^2 \, dr
\end{equation}

\noindent but at $r_{min}$ and $r_{max}$ we have $p_r = \partial
W_r /\partial r = 0$ and therefore the left hand side of the
equality is zero. The integrand in the right hand side is positive
and because $\dot{r}$ is not identical to zero, we have $r_{max}
\not= r_{min}$. Therefore, the integral becomes non-zero, and the
equality holds when we have $c_{\theta} = 0$. The value of
$c_\theta$ is a constant for the whole path of the particle.
Therefore, from \eqref{E:fr=const} we obtain for the whole path of
particle that

\begin{equation}\label{E:Lambda r}
    r^2 R_r^2 \frac{\partial W_r}{\partial r}=r^2 R_r^2 m
    \dot{r}=\lambda_r
\end{equation}

\noindent in which $\lambda_r$ is a constant. If $\lambda_r \not=
0$ we have $\dot{r} \not= 0$ for the whole path of the particle.
This means that, there is no turning point in the trajectory of
the particle. The particle either approaches the center of
potential or move away from it. This state can not be a bound one.
Therefore, for a bound system we must have $\lambda_r =
0$ and also identically $\dot{r} = 0$. Therefore,\\

\begin{quote}
"for conservative states of a bound system we \emph{always} have
$c_\theta = 0$ and $\dot{r}=0$".\\
\end{quote}

Now, consider the coordinate $\theta$ and let $c_\theta = 0$. When
we have identically $\dot{\theta}=0$, we get from Eq. \eqref{E:f
theta=const} that $c_\phi = 0$. If $\dot{\theta}$ is not
identically zero and system is bound, then there necessarily
exists at least one $\theta_{min}$ and one $\theta_{max}$ in the
path of particle. Putting $c_\theta = 0$ in Eq. \eqref{E:f
theta=const} and integrating from one $\theta_{min}$ to the
subsequent $\theta_{max}$, we obtain

\begin{equation}\label{E:f theta integration}
    (\sin \theta R_\theta^2 \frac{\partial W_\theta}{\partial
    \theta})_{\theta_{max}} - (\sin \theta R_\theta^2 \frac{\partial W_\theta}{\partial
    \theta})_{\theta_{min}} = - c_\phi \int_{\theta_{min}}^{\theta_{max}} \frac{R_\theta^2}{\sin \theta}
    \, d \theta.
\end{equation}

Similar to the previous reasoning for $c_\theta$, this equality
holds when $c_\phi = 0$. This means that we get from \eqref{E:f
theta=const}

\begin{equation}\label{E:Lambda theta}
    \sin \theta R_\theta^2 \frac{\partial W_\theta}{\partial
    \theta}=\sin \theta \, R_\theta^2 m r^2
    \dot{\theta}=\lambda_\theta
\end{equation}

\noindent in which $\lambda_\theta$ is a constant. Now, if
$\lambda_\theta \not= 0$, then $\dot{\theta}$ can never vanishes.
Thus, $\theta$ either increases or decreases, and since $\theta$
changes between $0$ and $\pi$, it reaches its limits and exceeds
them, but according to definition of $\theta$, this is impossible.
Therefore, the continuation of motion necessitates to have
$\dot{\theta}=0$ somewhere, which contradicts our assumption.
Thus, we must have $\lambda_\theta=0$, and since $\sin \theta$ or
$R_\theta$ are not identical to zero, we must always
have $\dot{\theta}=0$. Therefore\\

\begin{quote}
"for conservative states of a bound system we \emph{always} have
$c_\phi = 0$ and
$\dot{\theta}=0$".\\
\end{quote}

Putting $c_\phi = 0$ in \eqref{E:f phi=const} we get

\begin{equation}\label{E:Lambda phi}
    R_\phi^2 \frac{\partial W_\phi}{\partial \phi}=R_\phi^2 m r^2
    \sin^2 \theta \, \dot{\phi}=\lambda_\phi
\end{equation}

\noindent in which $\lambda_\phi$ is a constant.

We summarize the results of vanishing $c_\theta$, $c_\phi$,
$\dot{r}$ and $\dot{\theta}$ for stationary bound states. We
obtain from \eqref{E:Lz2 psi on psi -2} and \eqref{E:L2 psi on psi
-2} that the constants of motion $\alpha_\theta^2$ and
$\alpha_\phi^2$ are eigenvalues of operators $\hat{L}^2$ and
$\hat{L}_z^2$, respectively. Formerly, we observed from Eq.
\eqref{E:H psi on psi} that the other constant of motion i.e.
energy is eigenvalue of operator $\hat{H}$. Because of vanishing
$p_\theta$ and $p_r$, the Eqs. \eqref{E:p_theta2 Main Eq} and
\eqref{E:pr2 Main Eq} reduce to angular and radial parts of
Schr\"odinger equation.\\

There is no necessity to have $\dot{\phi} = 0$ for the bound
systems but it can occur. Indeed, there are two cases which are
consistent with Eqs. \eqref{E:p_phi2 Main Eq} and \eqref{E:Lambda
phi}. We can take $\dot{\phi} = 0$ (i.e. $p_\phi = 0$) or $R_\phi
= const$ (i.e. $p_\phi = \alpha_\phi$). In the first case, we
conclude from \eqref{E:p_phi2 Main Eq} that $R_\phi$ is an
eigenfunction of operator $\hat{L}_z^2$, and $\alpha_\phi$ take
values $m\hbar$ for integer $m$. In the second case we conclude
from \eqref{E:p_phi2 Main Eq} that $p_\phi = \alpha_\phi$ and
there is no a way to quantize $\alpha_\phi$ but imposing

\begin{equation}\label{E:W-S Quantum Condit.}
    \oint \nabla S. d x = integer \times 2\pi\hbar
\end{equation}

\noindent which is equivalent to appealing to the wave function
concept. According to Eq. \eqref{E:Lz psi on psi} the wave
function $\psi$ becomes eigenfunction of $\hat{L}_z$. The
condition \eqref{E:W-S Quantum Condit.} means that the $S$
function is unique apart from additive constants. This condition
allows us to introduce the single-valued wave function $\psi=R
\exp{(iS/\hbar)}$. Without condition \eqref{E:W-S Quantum Condit.}
we can not have a well-defined wave function.

Therefore, for quantized $\alpha_\phi$ we can either drop the wave
function or appeal to it. If you think in a quasi-Newtonian
framework, you can accept the first case, and if you think about
the complex wave function as a fundamental entity, you can accept
the second case.

\indent Consequently, one can explain the quantization of
$\alpha_\phi$ without appealing to the wave function. Of course,
we must remember that in this case $p_\phi$ and consequently
$\dot{\phi}$ are always zero. This with vanishing of $\dot{r}$ and
$\dot{\theta}$, leads to the conclusion that in the stationary
states of central potentials, the particle is always at rest. Thus
we can say that, contrary to what is stated in the articles on
ordinary Bohmian mechanics, the electron in hydrogen atom is at
rest not only when the magnetic quantum number $m$ is zero, but it
is at rest even for $m \neq 0$. In this case, the function $\psi$
is an eigenfunction of $\hat{H}$ and $\hat{L}^2$ and also of
$\hat{L}_z^2$, rather than $\hat{L}_z$ and becomes

\begin{equation*}\
\begin{aligned}
    \psi_e(r, \theta, \phi)&=R_{nl}(r)P_l^m(\cos \theta)\cos m
    \phi \\
    \psi_o(r, \theta, \phi)&=R_{nl}(r)P_l^m(\cos \theta)\sin m
    \phi \quad (m \neq 0)
\end{aligned}
\end{equation*}

\noindent rather than

\begin{equation*}
    \psi(r, \theta, \phi)=R_{nl}(r)P_l^m(\cos \theta)e^{im\phi}.
\end{equation*}

In the other words, assuming that $\psi$ is an eigenfunction of
$\hat{L}_z^2$ rather than $\hat{L}_z$, $\psi$ becomes real and we
reach the conclusion that the electron in all eigenfunctions of a central potential is at rest.\\

\section{Summary}\label{S:Summary}
\nin As we observed in this paper, one can solve the quantum
problems by Hamilton's canonical equations. We observed that
considering Hamilton's canonical equation along with the
continuity condition yield the quantization of energy and angular
momentum in a natural way without appealing to the 'eigenvalue
postulate'. This approach \emph{is a new kind of quantization},
based directly and completely on the Bohmian mechanics. The
presence of a non-trivial (Bohmian) quantum potential in
Hamilton's equations permits the existence of stable conservative
states $\partial R /
\partial t = 0$, and the presence of continuity condition compels the
energy and angular momentum for these states to be quantized. This
fact shows that the Bohmian mechanics is on a correct route. This
fact also shows the merit of writing Bohmian mechanics in the form
of modified Hamilton-Jacobi and continuity equations.

\indent According to this paper, the operator methods of ordinary
quantum mechanics are often useful, but we should not consider
them as the basis of theory of particle mechanics. We should
consider the operators and operator algebra as merely useful
mathematical tools for solving problems.\\
\indent Here we did not consider the Schr\"odinger equation as the
basis of our work, and we emphasized that the concept of wave
function is not necessarily a basic quantum concept, and that one
can solve quantum problems without appealing to it. But, this does
not mean that we are denying the practical value of the
Schr\"odinger equation. We can combine the real equations of
Bohmian mechanics in the form of complex Schr\"odinger equation
and use it to solve large number of problems. Indeed, as we
mentioned in previous paper, we must be aware that the form of
quantum potential is a mathematical necessity for minimizing the
total energy of ensemble (without referring to the wave function
and Schr\"odinger equation), and the quantization of energy and
angular momentum is a consequence of continuity condition for
stationary states.

\end{document}